\documentstyle[12pt,epsf]{article}

\begin{document}

\hoffset = -1truecm
\voffset = -2truecm
\baselineskip = 10 mm

\title{\bf Nuclear Shadowing and Antishadowing in a Unitarized BFKL Equation\thanks{Supported
by National Natural Science Foundation of China 10135060, 10475028
and 10205004.}}

\author{
{\bf RUAN Jian-Hong}, {\bf SHEN Zhen-Qi} and {\bf ZHU
Wei\footnote{Corresponding author, E-mail: weizhu
@mail.ecnu.edu.cn}}\\
\normalsize (Department of Physics, East China Normal University,
Shanghai 200062, China) \\
}

\date{}

\newpage

\maketitle

\vskip 3truecm

\begin{abstract}

    The nuclear shadowing and antishadowing effects are explained by a unitarized
BFKL equation. The $Q^2$- and $x$-variations of the nuclear parton
distributions are detailed based on the level of the unintegrated
gluon distribution. In particular, the asymptotical behavior of the
unintegrated gluon distribution near the saturation limit in nuclear
targets is studied. Our results in the nuclear targets are
insensitive to the input distributions if the parameters are fixed
by the data of a free proton.

\end{abstract}


$keywords$:  nuclear shadowing and antishadowing; unintegrated gluon
distribution in nuclei; QCD evolution equation

\newpage
\begin{center}
\section{Introduction}
\end{center}

    The gluon distribution in the nuclear target is
an essential ingredient in the calculation of high energy nuclear
collisions, which relate to minijet production, dilepton production,
heavy quarks and their bound states. Although the gluon density in a
free nucleon was extracted by the experiments, however there are not
enough data for the gluon distributions in nuclei. As well known the
parton densities in a bound nucleon differ from in a free nucleon,
the ratio of nuclear and deuterium structure functions is smaller or
lager than unity at Bjorken variable $x<0.05-0.1$ or $x\simeq
0.1-0.2$. These two facts are called the nuclear shadowing and
antishadowing effects, respectively [1]. The understanding of
nuclear shadowing and antishadowing in QCD is therefore an important
issue to predict the nuclear gluon distributions. In this aspect,
the parton recombination (fusion) between two different nucleons in
a nucleus is a natural mechanism, which transfers the partons from a
smaller $x$ region to a larger $x$ region and forms the nuclear
shadowing and antishadowing effects [2].

     The shadowing and antishadowing phenomena are also predicted to
happen in a free nucleon: a rapid rise of parton multiplicities
inside the proton at small $x$ leads to the gluon recombination,
which changes the distributions of gluon- and quark-densities, but
does not change their total momenta. In consequence, a part of the
gluon momentum lost in the shadowing should be compensated in terms
of new gluons with larger $x$, which form the antishadowing. The
modification of the gluon recombination to the standard DGLAP [3]
evolution equation was first proposed by Gribov-Levin-Ryskin and
Mueller-Qiu (the GLR-MQ equation) in [4], and it is naturally
regarded as the QCD dynamics of the nuclear shadowing since the type
of gluon fusion in a nucleon is still there in a nucleus but roughly
differs by an $A^{1/3}$ scale [5]. However, the GLR-MQ equation does
not predict the nuclear antishadowing effect since the momentum
conservation in the gluon recombination is violated in this
equation. For this sake, a modified DGLAP equation was proposed to
replace the GLR-MQ equation in [6,7], where the corrections of the
gluon fusion to the DGLAP equation lead to the shadowing and
antishadowing effects. Unfortunately, the integral solutions either
in the GLR-MQ equation or in other modified DGLAP equations need the
initial distributions on a boundary line $(x,Q^2_0)$ at fixed
$Q^2_0$, and they contain the unknown input nuclear shadowing and
antishadowing effects. Close, Qiu and Roberts [8] constructed a
parton fusion model by QCD arguments and try to evolve the input
parton distributions in the nuclear target, however, the results of
this parton fusion model in the small $x$ region are sensitive to
the input parton distributions [9], which leads to a large
uncertainty in the predictions. On the other hand, recently nuclear
unintegrated gluon distribution becomes a useful and intuitive
phenomenological language for applications to many high-energy
nuclear collisions. However, the DGLAP equation and its modified
forms are based on the collinear factorization scheme and they don't
predict the evolution of the unintegrated parton distributions.

    Instead of the above-mentioned modified DGLAP equations,
an alternative QCD research for the nuclear shadowing and
antishadowing is to modify the BFKL equation [10], which is directly
written for the unintegrated gluon distribution. The evolutions of
the BFKL-type equations are along the small $x$-direction: the input
gluons distribute on the boundary line $(x_0,\underline{k}^2)$ at
fixed $x_0$, where the gluon fusion begins, and $all$ parton
distributions both in a free nucleon and in a bound nucleon at
$x<x_0$ are the evolution results of these equations. There are
several nonlinear evolution equations considering the corrections of
the gluon fusion to the BFKL evolution. One of the most widely
studied models is the Balitsky-Kovchegov (BK) equation [11]. The BK
equation is originally derived for the scattering amplitude in the
transverse coordinator space. The nonlinear terms in the BK equation
are formed by the dipole splitting and the screening effect origins
from the double scattering of the probe on the target. A remarkable
solution of the BK equation is the so-called saturation, where the
amplitude is a completely flat spectrum. However, the amplitude in
the BK equation is re-normalized to identify the coefficient of the
nonlinear term with that of the linear term. Therefore, the relation
of the unintegrated gluon distribution with this scattering
amplitude is unclear and it is model-dependent [12]. Besides, it is
similar to the GLR-MQ equation, the BK equation is irrelevant to the
antishadowing effect. Recently, a new modified BFKL equation
incorporating the shadowing and antishadowing corrections of the
gluon recombination to the unintegrated gluon distribution was
proposed by Ruan, Shen, Yang and Zhu (the RSYZ equation) [13].

    The purpose of this work is to explain the observed nuclear
shadowing and antishadowing effects for the parton distributions
using the RSYZ equation. The $Q^2$- and $x$-variations of the
nuclear parton distributions in the RSYZ equation are detailed in
this work. Particularly, we predict the unintegrated nuclear gluon
distributions using the RSYZ equation. The results show a
logarithmic increasing spectral height, which is not identical to
the prediction of the BK equation in the transverse coordinator
space but is similar to a mean field result.

    The paper is organized as follows. In Section 2 we give the
RSYZ equation and its modifications in the nuclear target.  The
predictions of the RSYZ equation to the nuclear shadowing and
antishadowing effects in the quark- and gluon-distributions as
well as in the unintegrated gluon distributions are presented in
Section 3.

\newpage
\begin{center}
\section{The RSYZ evolution equation}
\end{center}

    The unintegrated gluon distribution $f_N$ in a proton
obeys the RSYZ evolution equation at small $x$ [13], where the
contributions of the gluon recombination to the BFKL dynamics are
considered

$$-x\frac{\partial f_N(x,\underline{k}^2)}{\partial x}$$
$$=\frac{\alpha_{s}N_c\underline{k}^2}{\pi}\int_{\underline{k}'^2_{min}}^{\infty} \frac{d \underline{k
}'^2}{\underline{k}'^2}\left\{\frac{f_N(x,\underline{k}'^2)-f_N(x,\underline{k}^2)}
{\vert
\underline{k}'^2-\underline{k}^2\vert}+\frac{f_N(x,\underline{k}^2)}{\sqrt{\underline{k}^4+4\underline{k}'^4}}\right\}$$
$$-\frac{36\alpha_s^2}{\pi\underline{k}^2R^2}\frac{N_c^2}{N_c^2-1}
f_N^2(x,\underline{k}^2)+\frac{18\alpha_s^2}{\pi\underline{k}^2R^2}\frac{N_c^2}{N_c^2-1}
f_N^2\left(\frac{x}{2},\underline{k}^2\right),~~x\le x_0;$$

$$-x\frac{\partial f_N(x,\underline{k}^2)}{\partial x}$$
$$=\frac{\alpha_{s}N_c\underline{k}^2}{\pi}\int_{\underline{k}'^2_{min}}^{\infty} \frac{d \underline{k
}'^2}{\underline{k}'^2}\left\{\frac{f_N(x,\underline{k}'^2)-f_N(x,\underline{k}^2)}
{\vert
\underline{k}'^2-\underline{k}^2\vert}+\frac{f_N(x,\underline{k}^2)}{\sqrt{\underline{k}^4+4\underline{k}'^4}}\right\}$$
$$+\frac{18\alpha_s^2}{\pi\underline{k}^2R^2}\frac{N_c^2}{N_c^2-1}
f_N^2\left(\frac{x}{2},\underline{k}^2\right),~~x_0\le x\le 2x_0,
\eqno(1)$$ where $R=1~fm$ (or $< 1~fm$) if the gluons are
uniformly distributed in a nucleon (or the gluons are located in
the hot-spots); $x_0$ is the starting point of the gluon fusions.
Note that the shadowing and antishadowing coexist in the region
$x\le x_0$, while there is only the antishadowing in $x_0\le x\le
2x_0$ [14].

    We should point out that the nonlinear terms in Eq. (1) are
really from the evolution kernel of the MD-DGLAP equation [6],
where the double logarithmic approximation is taken, i.e., both
the $x$ and transverse momenta are strongly ordered. Obviously,
this approximation satisfies such $x$-region, where the values of
$x$ are not extra small. On the other hand, the linear terms of
Eq. (1) are the BFKL-kernel, which works in the small $x$ range.
The mix of two approximations in Eq. (1) is feasible for the
discussions of the nuclear shadowing and antishadowing effects,
since they occupy the transition range from a middle $x$ to small
$x$. The RSYZ is directly derived for the unintegrated gluon
distribution, which relates to the (integrated) gluon distribution
using

$$G_N(x,Q^2)\equiv
xg_N(x,Q^2)=\int^{Q^2}_{\underline{k}^2_{min}}\frac{d\underline{k}^2}{\underline{k}^2}
f_N(x,\underline{k}^2). \eqno(2)$$

    Since the density of gluons increases rapidly with decreasing
    $x$,
the sea quark distributions are increasingly dominated by the
gluon distribution, via the DGLAP splitting $G\rightarrow
q\overline{q}$. Thus, the deep inelastic structure function at
small $x$ reads [15]

    $$F_{2N}(x,Q^2)$$
$$=2\int^1_xdx_1\int^{Q^2}\frac{d\underline{k}^2}{\underline{k}^2}
\int^{\underline{k}^2}\frac{d\underline{k}'^2}{\underline{k}'^2}f_N(\frac{x}{x_1},\underline{k}'^2)
\sum_qe^2_q\frac{\alpha_s}{2\pi}P_{qG}(x_1). \eqno(3)$$ where
$P_{qG}(x_1)$ is the DGLAP splitting function.

    Now we discuss the RSYZ equation in the nuclear target. The gluons with smaller $x$ exceed the longitudinal
size of nucleon in a nucleus.  We assume that the gluons inside the
nucleus are completely overlapping and fusion along the longitudinal
momentum direction at the evolution starting point $x_0$. Thus, the
strength of the nonlinear recombination terms in Eq. (1) should be
scaled by $A^{1/3}$ in a nucleus. On the other hand, although the
softer gluons of different nucleons with extra small
$\underline{k}^2$ may be correlated on the transverse plane because
the integrations on $\underline{k}^2$ can go down to a very small
value in Eq. (1), we neglect these corrections due to
$f_N(x,\underline{k}^2\rightarrow 0)\rightarrow 0$.  In this simple
model the RSYZ equation in the nucleus becomes

$$-x\frac{\partial f_A(x,\underline{k}^2)}{\partial x}$$
$$=\frac{\alpha_{s}N_c\underline{k}^2}{\pi}\int_{\underline{k}'^2_{min}}^{\infty} \frac{d \underline{k
}'^2}{\underline{k}'^2}\left\{\frac{f_A(x,\underline{k}'^2)-f_A(x,\underline{k}^2)}
{\vert
\underline{k}'^2-\underline{k}^2\vert}+\frac{f_A(x,\underline{k}^2)}{\sqrt{\underline{k}^4+4\underline{k}'^4}}\right\}$$
$$-A^{1/3}\frac{36\alpha_s^2}{\pi\underline{k}^2R^2}\frac{N_c^2}{N_c^2-1}
f^2_A(x,\underline{k}^2)+A^{1/3}\frac{18\alpha_s^2}{\pi\underline{k}^2R^2}\frac{N_c^2}{N_c^2-1}
f^2_A\left(\frac{x}{2},\underline{k}^2\right),~~x\le x_0;$$

$$-x\frac{\partial f_A(x,\underline{k}^2)}{\partial x}$$
$$=\frac{\alpha_{s}N_c\underline{k}^2}{\pi}\int_{\underline{k}'^2_{min}}^{\infty} \frac{d \underline{k
}'^2}{\underline{k}'^2}\left\{\frac{f_A(x,\underline{k}'^2)-f_A(x,\underline{k}^2)}
{\vert
\underline{k}'^2-\underline{k}^2\vert}+\frac{f_A(x,\underline{k}^2)}{\sqrt{\underline{k}^4+4\underline{k}'^4}}\right\}$$
$$+A^{1/3}\frac{18\alpha_s^2}{\pi\underline{k}^2R^2}\frac{N_c^2}{N_c^2-1}
f^2_A\left(\frac{x}{2},\underline{k}^2\right),~~x_0\le x\le 2x_0.
\eqno(4)$$

    The distributions $G_A(x,Q^2)$ and $F_{2A}(x,Q^2)$ in the nucleus are
computed by using the equations corresponding to Eqs. (2) and (3).

\newpage
\begin{center}
\section{Numerical analysis and summary}
\end{center}

    We use a parameter form of a BFKL-like solution as the input
distribution [13] at $2x_0=0.3$

$$
f_N(2x_0,\underline{k}^2)=f_A(2x_0,\underline{k}^2)=\beta\sqrt{\underline{k}^2}\frac{x_0^{-\lambda_{BFKL}}}{\sqrt
{\ln{\frac{1}{x_0}}}}exp\left(-\frac{\ln^2(\underline{k}^2/1~GeV^2)}{2\lambda'\ln
(1/x_0)}\right), \eqno(5)$$ where $\lambda_{BFKL}=12\alpha_s/(\pi
\ln 2)$ and $\beta$ and $\lambda'$ are two parameters.

    The computations of the RSYZ equation need pre-know the value
of $f_{N(A)}(x_i/2,\underline{k}^2)$ at the step $x=x_i$. For this
sake, we proposed the following program in [13]

$$f_{N(A)}\left(\frac{x_i}{2},\underline{k}^2\right)=f_{N(A),Shadowing}\left(\frac{x_i}{2},\underline{k}^2\right)
+\frac{f_{N(A),BFKL}\left(\frac{x_i}{2},\underline{k}^2\right)-f_{N(A),Shadowing}\left(\frac{x_i}{2},\underline{k}^2\right)}{i\eta
-\eta+1} ,\eqno(6)$$ where
$f_{N(A),Shadowing}(x_i/2,\underline{k}^2)$ (or
$f_{N(A),BFKL}(x_i/2,\underline{k}^2)$) indicates that the evolution
from $x_i$ to $x_i/2$ is controlled by Eq. (1) but without the
antishadowing contributions (or is controlled by the BFKL equation).
The parameter $\eta$ implies the different velocities approaching
the BK dynamics. We temporarily take $\eta=\infty$ and we will
indicate it is appropriate.

       At first, we use the well known $F_{2N}(x,Q^2)$-data [16] of a free proton to
determine the parameters in the computations. Then we predict the
distributions in nuclei. In this work we fix the coupling constant
to be $\alpha_s=0.3$. The dashed curve in Fig. 1 is our fitting
result using $\beta=7.22$, $\lambda'=0.002$ and $R=2.6~GeV^{-1}$.
Note that the contributions from the valence quarks to $F_2$ at
$x>0.1$ are necessary and they can be parameterized by the
difference between the dashed curve and the experimental solid curve
in Fig. 1.

    Figure 2 shows the predictions of the RSYZ equation for the
Ca/C, Ca/Li, Ca/D and Cu/D compared with the EMC and NMC results
[17,18]. The agreement is acceptable. Figure 3 indicates that the
enhancement in the Sn gluon distribution with respect to that in C
observed by NMC [19] is consistent with our predictions.

    No significant $Q^2$-dependence on the ratios of the structure
functions at small $x$ has been concluded in the present
experimental precision. However, it does not prevent us exposing
the possible $Q^2$-variations of the nuclear shadowing, which may
be hidden in the larger experimental errors. The ratios
$G_{Ca}(x,Q^2)/G_D(x,Q^2)$ for gluon distributions at $Q^2=2$ and
$10~GeV^2$ using Eqs. (1) and (2) are given in Fig.4a. The
$Q^2$-variations of the gluon ratios are predicted in the region
$10^{-4}<x<10^{-1}$ in our model. The logarithmic slope $b$ in
$G_A/G_{A'}=a+b\ln Q^2$ is positive. However, the corresponding
slope in the ratio of the structure functions $F_{2Ca}/F_{2D}$ is
negative (see Fig.4b). For example, the predicted $Q^2$-slope for
calcium at $x\simeq 4\times10^{-2}$, $b\simeq -0.046$, and at
$x\simeq 10^{-2}$, $b\simeq -0.03$, the results are compatible
with the measured data in [20]. A more significant
$Q^2$-dependence of the structure function ratios can be found in
the heavy nucleus $F_{2Xe}/F_{2D}$. A flatter ratio at $x<10^{-2}$
in the E665 data [21] was presented. However, it can't be
understood as the saturation behavior, where the parton fusion
balances the parton splitting. The reason is that the parton
densities in the proton still increase toward the small
$x$-direction at $x<0.01$ (see Fig.1). Alternatively, we consider
that this behavior is a consequence of the $Q^2$-variations  of
the structure function ratios since the experimental point with
the smaller $x$ corresponds to the smaller value of $Q^2$ in Fig.
5.

        An important parameter in the computing RSYZ equation is $\eta$ in Eq. (5).
The observed nuclear shadowing and antishadowing provide an
example to determine the value of $\eta$. The results incline to a
minimum antishadowing, i. e., $\eta=\infty$.

        Comparing with the modified DGLAP equation, the RSYZ equation
directly predicts the nuclear unintegrared gluon distributions,
which are important information for the researches of high energy
nuclear collisions. We compute the unintegrated gluon
distributions in different nuclear targets using the RSYZ
equation. In Fig. 6 we plot the spectra of the gluon density on
the transverse momentum squared $dG/d\underline{k}^2\equiv
dG/dQ^2\vert_{Q^2=\underline{k}^2}=f(x,\underline{k}^2)/\underline{k}^2\sim
\underline{k}^2$. The BK equation on the transverse coordinator
space predict the saturation solutions, where the amplitude is a
completely flat spectrum at $\underline{k}^2<Q_s^2(x)$, and the
spectral height is irrelevant to the values of $x$ (i.e., the
geometric scaling); $Q^2_s(x)$ is the saturation scale. The
saturation limit and the geometric scaling are expected to appear
in the larger nucleus, where the shadowing effects are enhanced.
However, the logarithmic increasing spectral height in our results
is not identical to this prediction of the BK equation but is
similar to the mean field result: $dG/d\underline{k}^2\propto \ln(
Q^2_s(x)/\underline{k}^2)$ [22].

     Finally, we discuss the dependence of our solutions with the input
distributions. For this sake, we change the parameters in Eq. (24)
to: $\beta=0.15$, $\lambda'=9.64$, $R=2.4~GeV^{-1}$ and strongly
distort the input form. Then we repeat our calculating programs.
We find that the results  are insensitive to the input
distributions.

     In summary, the nuclear shadowing and antishadowing effects
are explained by a unitarized BFKL equation. The $Q^2$- and
$x$-variations of the nuclear parton distributions are detailed
based on the level of the unintegrated gluon distribution. In
particular, the asymptotical behavior of the unintegrated gluon
distribution in various nuclear targets is studied. We find that the
geometric scaling in the expected saturation range is violated. Our
results in the nuclear targets are insensitive to the input
distributions if the parameters are fixed by the data of a free
proton. We believe that these results are useful in studying the
ultrarelastivistic heavy ion collisions.

\newpage

Figure Captions

Fig. 1  The fit of the computed $F_{2P}(x,Q^2=10~GeV^2)$ in proton
by the RSYZ equation using the input Eq. (5) (dashed curve). The
contributions of the valence quarks are parameterized by the
differences between solid and dashed curves. The data are taken
from [16].

Fig. 2  Predictions of the RSYZ equation compared with the ratios
of the structure functions for various nuclei. The data are taken
from [17,18]. All curves are for $Q^2=10~GeV^2$

Fig. 3  Predictions for the ratio of the gluon distributions in
$Sn/C$. The curve is the result of the RSYZ equation for
$Q^2=10~GeV^2$ and the data are taken from [19].

Fig. 4  (a) $Q^2$-dependence of the ratios for the gluon
distributions $Ca/D$ in the RSYZ equation. (b) Similar to (a) but
for the the structure functions.

Fig. 5  $Q^2$-dependence of the ratios for the structure functions
$Xe/D$ in the RSYZ equation. The data are taken from [21].

Fig. 6  Predicted spectrum
$dG/d\underline{k}^2=f(x,\underline{k}^2)/\underline{k}^2$ on the
$\underline{k}^2$-space in various nuclear targets in the RSYZ
equation. Dotted curves $x=10^{-5}$, dashed curves $x=10^{-6}$ and
solid curves $x=10^{-7}$.

\newpage
\epsfxsize=15cm \epsffile{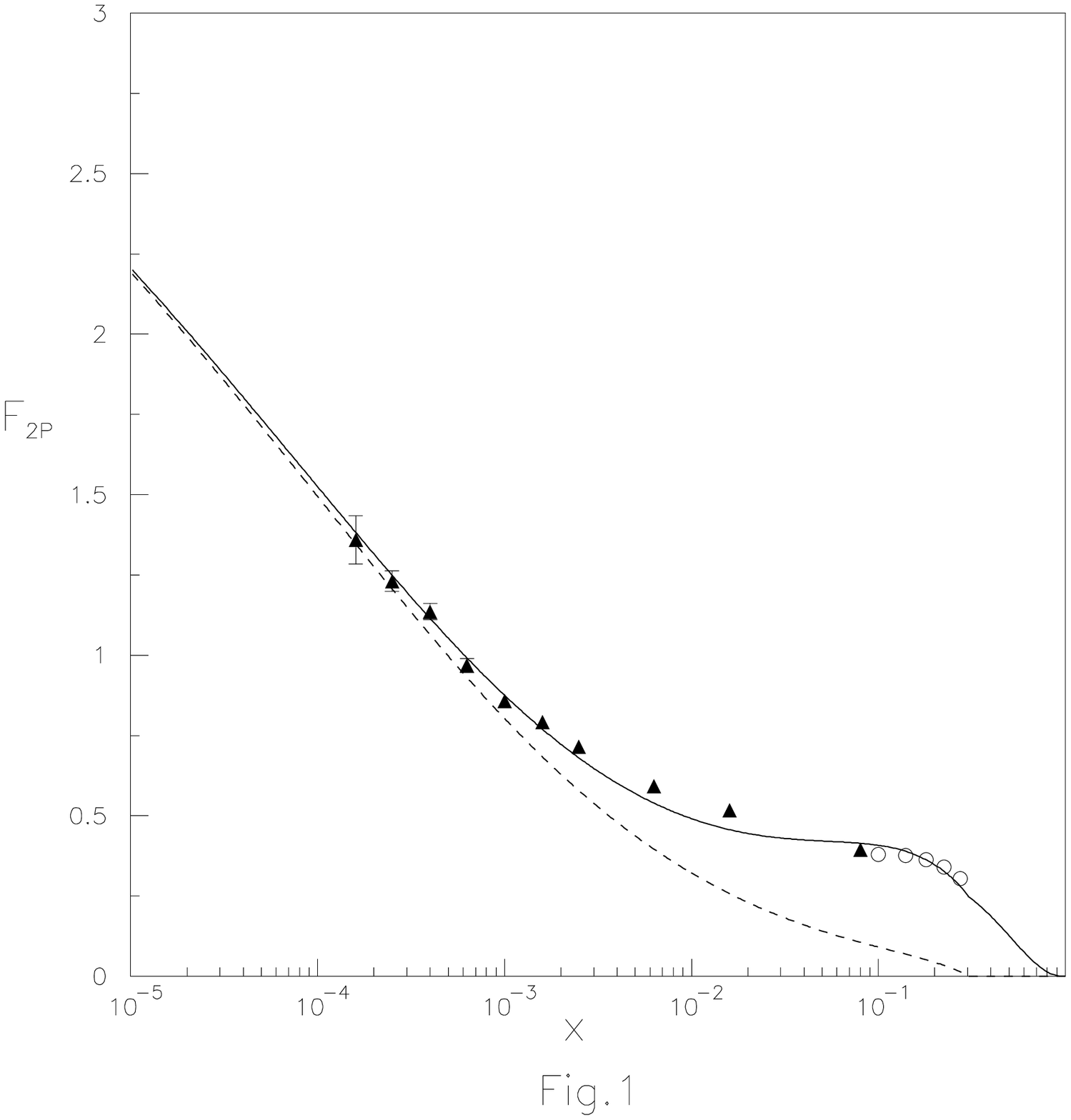}
\newpage
\epsfxsize=15cm \epsffile{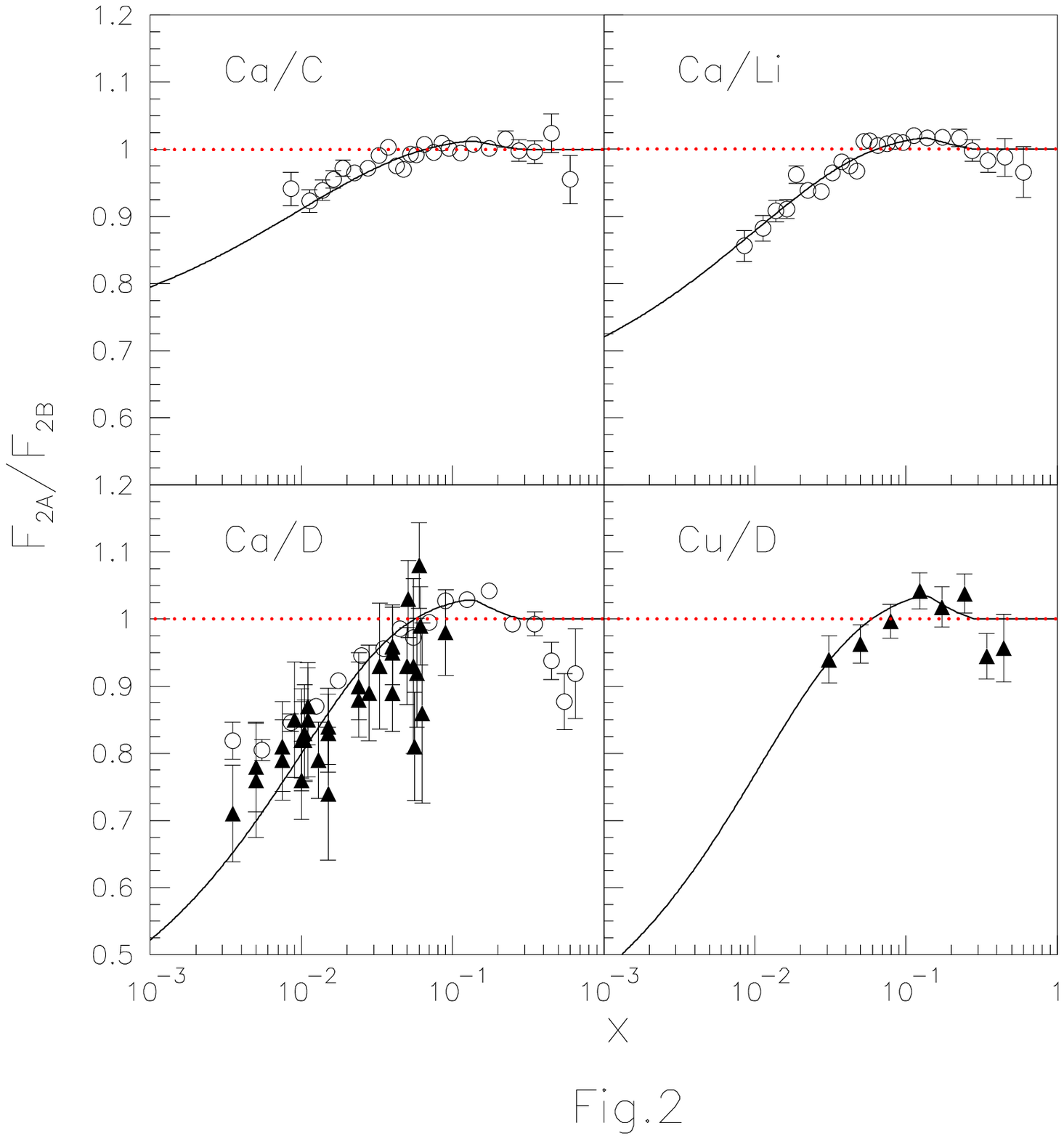}
\newpage
\epsfxsize=15cm \epsffile{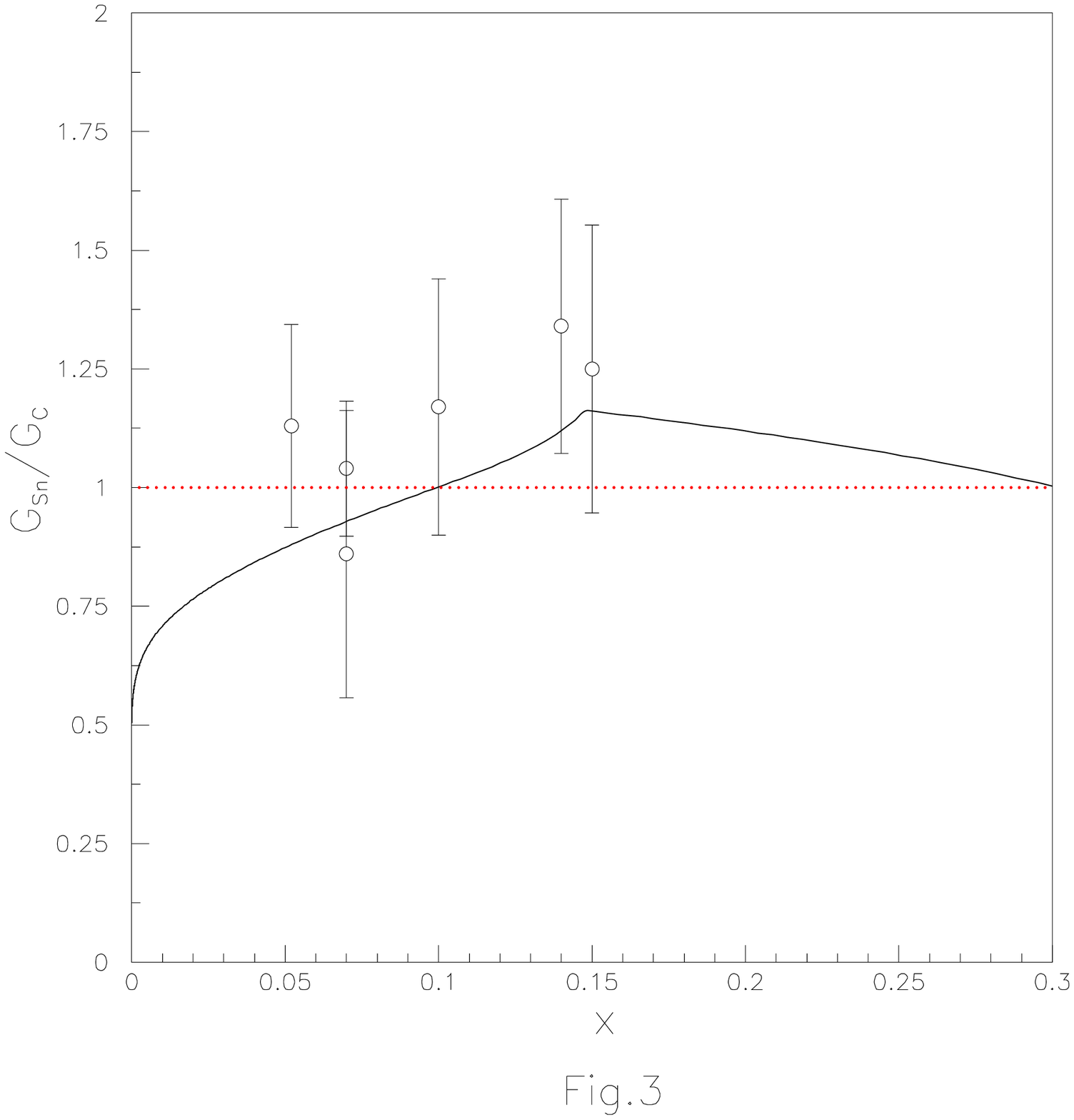}
\newpage
\epsfxsize=15cm \epsffile{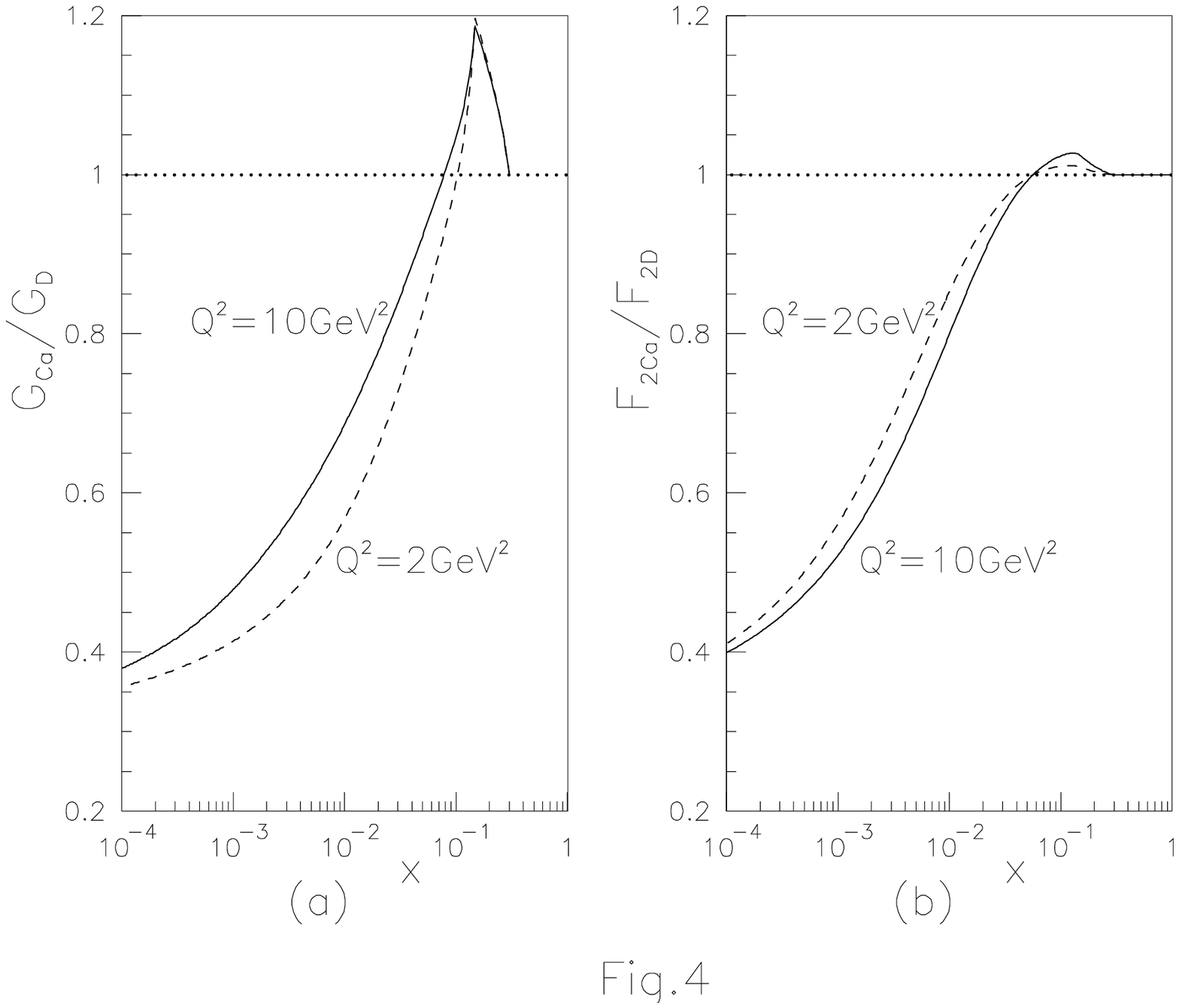}
\newpage
\epsfxsize=15cm \epsffile{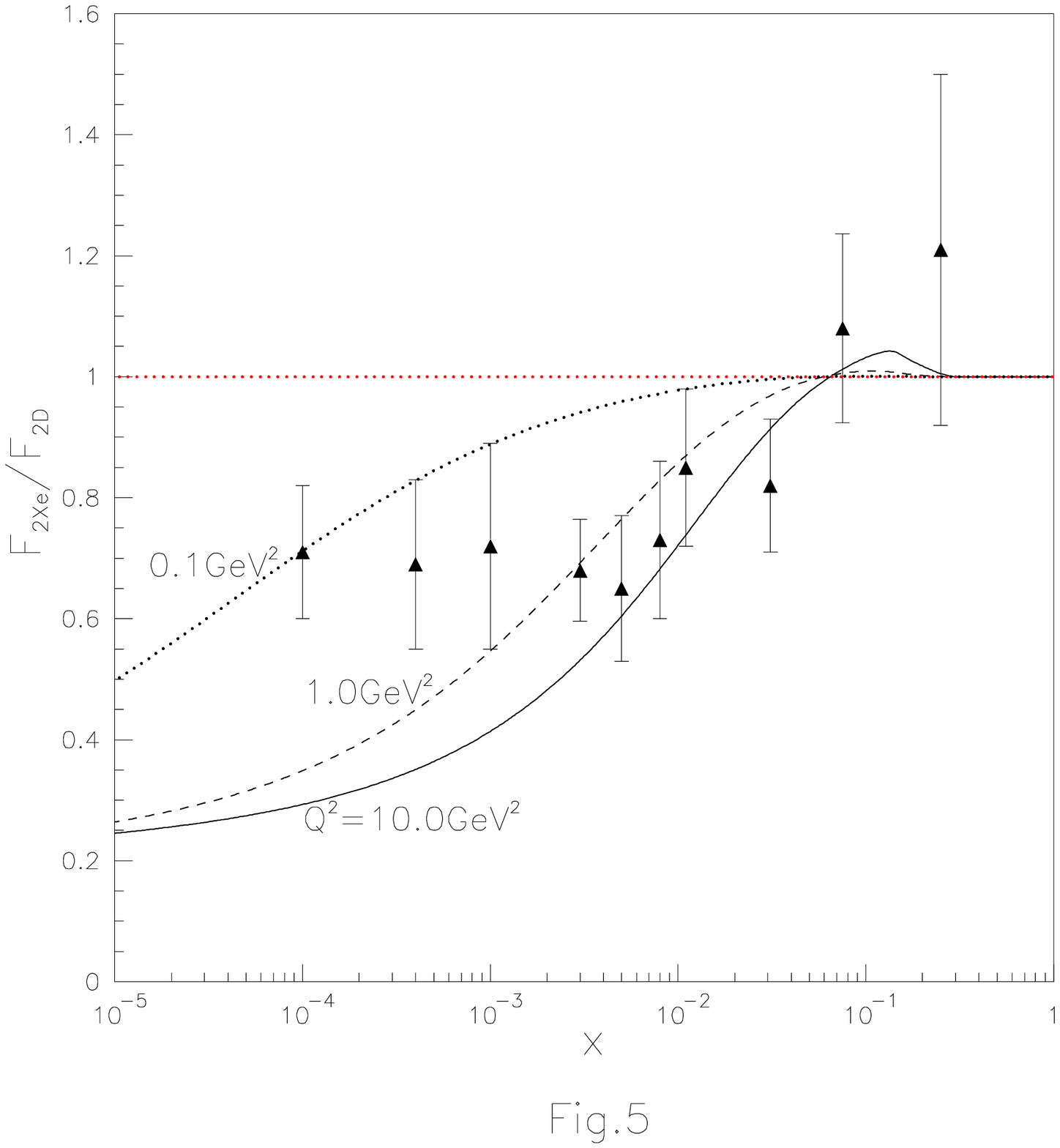}
\newpage
\epsfxsize=15cm \epsffile{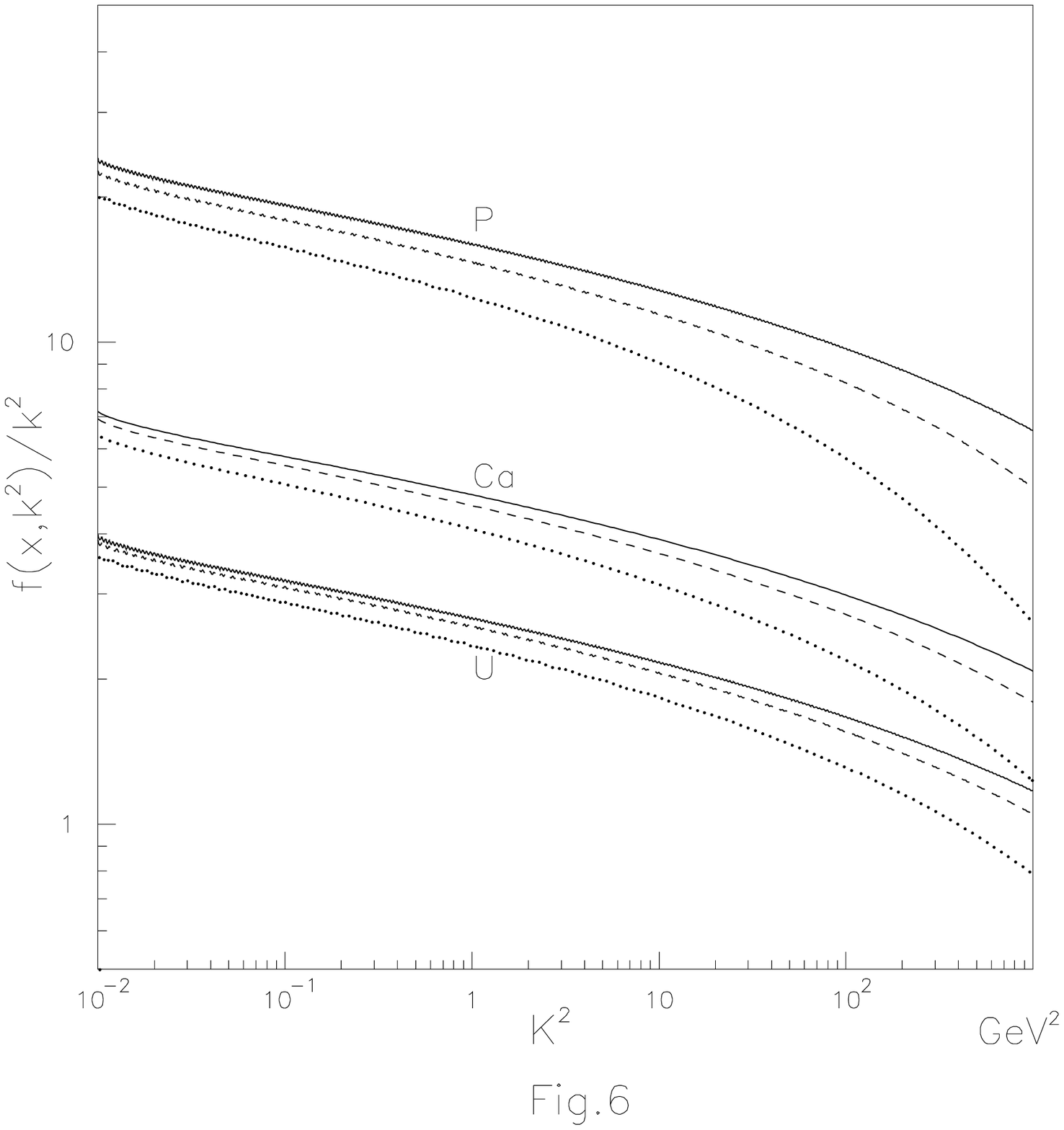}

\vspace{0.3cm}

\newpage
\begin{thebibliography}{99}

\bibitem{1} Arneodo M. Phys. Rept., 1994, 240: 301-393

\bibitem{2} Nikolaev N N, Zakharov V I. Phys. Lett., 1975,
B55:397-399


\bibitem{3} Altarelli G, Parisi G. Nucl. Phys., 1977, B126: 298-318;
Gribov V N, Lipatov L N. Sov. J. Nucl. Phys., 1972, 15: 438-450;
Dokshitzer Y L. Sov. Phys. JETP., 1977, 46: 641-653

\bibitem{4} Gribov L V, Levin E M and Ryskin M G. Phys. Rept.,
1983, 100: 1-150; Mueller A H, Qiu J. Nucl. Phys., 1986, B268:
427-452

\bibitem{5} Qiu J W. Nucl. Phys., 1987, B291: 746-764; Eskola K J,
Qiu J W and Wang X N. Phys. Rev. Lett., 1994, 72: 36-39; Eskola K
J,
 Honkanen H,  Kolhinen V J, Qiu J W and Salgado C A. Nucl. Phys.,
 2003, B660: 211-224

\bibitem{6}  ZHU Wei. Nucl. Phys., 1999, B551: 245-274

\bibitem{7} ZHU Wei, RUAN Jian-Hong, Yang Ji-Feng and SHEN Zhen-Qi, Phys.
Rev., 2003, D68: 094015.

\bibitem{8} Close F E, Qiu J W and  Roberts R G. Phys. Rev., 1989,
D40: 2820-2831.

\bibitem{9} Kumano S, Phys.Rev., 1993, C48: 2016-; Phys.Lett., 1993, B298:
171-175

\bibitem{10} Lipatov L N. Sov. J. Nucl. Phys., 1976, 23: 338-345; Fadin V S, Kuraev E.A.
and Lipatov L N. Phys. Lett., 1975, B60: 50-52; Kuraev E A,
Lipatov L N and Fadin V S, Sov. Phys. JETP., 1976, 44: 443-450;
Kuraev E A, Lipatov L N  and Fadin V S. Sov. Phys. JETP., 1977,
45: 199-204; Balitsky I I, Lipatov L N. Sov. J. Nucl. Phys., 1978,
28: 822-829

\bibitem{11} Balitsky I. Nucl. Phys., 1996, B463:99-; Kovchegov Yu.
Phys. Rev., 1999, D60:034008; Phys. Rev., 2001, D61:074018.

\bibitem{12} Kutak K, Stasto A M. Eur.Phys.J., 2005, C41: 343-351.

\bibitem{13} RUAN Jian-Hong, SHEN Zhen-Qi, YANG Ji-Feng and Zhu Wei. Nucl.
Phys., 2007, B760: 128-144.

\bibitem{14} ZHU Wei, XUE Da-Li, CHAI Kang-Ming and XU Zai-Xin. Phys.
Lett., 1993, B317: 200-204; ZHU Wei, CHAI Kang-Ming and HE Bo.
Nucl. Phys., 1994, B427: 525-533; ZHU Wei, CHAI Kang-Ming and HE
Bo. Nucl. Phys., 1995, B449: 183-196

\bibitem{15}  Askew A J, Kwiecinski J, Martin A D and
Sutton P J. Phys. Rev., 1993, D47: 3775-3782

\bibitem{16} Derrick M et al. Zeit.Phys., 1996, C72: 399-424; Benvenuti A C et
al. Phys. Lett., 1989, B223: 485-489

\bibitem{17} CERN NA28/EMC, Arneodo M et al. Phys. Lett., 1988,
B211: 493-499 ; Nucl. Phys., 1990, B333: 1-47

\bibitem{18} CERN NA37/NMC, Amaudruz P et. al. Nucl. Phys., 1995, B441:
3-11

\bibitem{19} CERN NA37/NMC, Amaudruz P et. al., Nucl. Phys., 1992,
B371: 553-566

\bibitem{20} Bodek A (SLAC E139), talk at the Lepton-Photon
Symposium and Europhysics Conference on High Energy Physics
LP-HEP91, Geneva, Switzerland, 25th July-1st August 1991.

\bibitem{21} FNAL E665, Adams M R et al. Phys. Rev. Lett., 1992,
68: 3266-3269

\bibitem{22} Iancu E, A. Leonidov, and L. McLerran, hep-ph/0202270, and
references therein.

\end{thebibliography}
\end{document}